\documentclass[a4paper,12pt]{article}
\usepackage{amsmath,amssymb}
\usepackage{bm}
\usepackage[dvipdfmx]{color,graphicx}
\usepackage{subfigure}
\usepackage{verbatim}
\usepackage{wrapfig}
\usepackage{ascmac}
\usepackage{makeidx}
\usepackage{url}
\usepackage{fancybox}  
\usepackage{amsfonts}
\usepackage{booktabs}
\usepackage{indentfirst}
\usepackage[small, bf]{caption}
\definecolor {gray} {gray} {0.3} 
\usepackage{cite}
\usepackage{feynmp}
\newcommand{\Slash}[1]{{\ooalign{\hfil#1\hfil\crcr\raise.167ex\hbox{/}}}}
\newcommand{\non}{\nonumber \\ }
\newcommand{\beq}{\begin{eqnarray}}
\newcommand{\eeq}{\end{eqnarray}}
\def\thefootnote{\ifnum\c@footnote>\z@\textasteriskcentered\@arabic\c@footnote\fi}
\makeatletter
\renewcommand{\footnoterule}{%
\kern-3\p@
\hrule width 0.4\columnwidth
\kern 2.6\p@}
\def\thefootnote{\ifnum\c@footnote>\z@\@arabic\c@footnote\fi}
\makeatother

\newcommand{\un}[1]{{\mathrm{\,#1}}}
\newcommand{\TeV}{\un{TeV}}
\newcommand{\GeV}{\un{GeV}}

\begin{document}

\begin{center}

\hfill UT-13-06\\ 
\hfill March, 2013\\

\vspace*{0.8cm}

\Large{\bf{Stau with Large Mass Difference and  Enhancement of  \bm{$h \rightarrow \gamma \gamma $ } Decay Rate in the MSSM\\ \vspace*{1.5cm}}}

\large{\textrm{\bf  Teppei Kitahara}}${}^\dag$ \footnote[0]{${}^\dag$ Electronic address: kitahara@hep-th.phys.s.u-tokyo.ac.jp }\large{\textrm{and \bf  Takahiro Yoshinaga}}${}^\ddagger$ \footnote[0]{${}^\ddagger$ Electronic address: t.yoshinaga@hep-th.phys.s.u-tokyo.ac.jp} \vspace*{0.8cm}

\textit{Department of Physics, The University of Tokyo, \\  \vspace{2mm} Tokyo 113-0033, Japan} \vspace*{1.6cm}

\end{center}

\begin{abstract}
The ATLAS and the CMS collaborations have   presented results which show an excess of the $h\rightarrow \gamma \gamma$ decay channel. 
In the Minimal Supersymmetric Standard Model (MSSM), this situation can be achieved by a light stau and a large left-right mixing of the staus.
 However, this parameter region is severely constrained by vacuum stability.  
In order to relax the vacuum meta-stability condition, we focus on the parameter region where  the mass difference between the two staus is large. 
This region has not been considered yet.
In this paper, we show that  staus with a large mass difference can relax the vacuum meta-stability condition sufficiently even if the lighter stau mass $m_{\tilde{\tau}_1}$ is kept light. 
We find that when the mass difference of two staus is large, the enhancement of the $h \rightarrow \gamma \gamma $ decay rate becomes small in spite of a relaxation of the vacuum meta-stability condition. Because of this feature,  an $\mathcal{O}(70)$\% enhancement  of $\Gamma(h\rightarrow \gamma \gamma )/\Gamma(h\rightarrow \gamma \gamma )_{\rm{SM}}$  is difficult to achieve in the  light stau scenario in the MSSM. 
\end{abstract}
{\small \hspace{21pt}  \textsc{ Keywords:} Supersymmetry, Higgs to diphoton decay rate, Vacuum stability}
\thispagestyle{empty}
\newpage
\section{Introduction}

The ATLAS \cite{:2012gk}  and the CMS \cite{:2012gu}  collaborations discovered  the Higgs-like particle, and measurements  of the Higgs couplings to Standard Model (SM) particles have started at the LHC. 
Especially, the measurement of the Higgs coupling to diphoton is interesting, 
since the Higgs to diphoton decay process is a one-loop level process in the SM and can be significantly influenced by new physics beyond the SM. 
Both the ATLAS and the CMS collaborations have reported that the observed diphoton signal strength is $1.5 - 1.8$ times larger than its SM prediction value, respectively \cite{ATLAS-CONF-2012-168, CMS-PAS-HIG-12-045}, 
\beq
\mu(\gamma \gamma )_\textrm{ATLAS}& = &1.80 \pm 0.30 \textrm{ (stat)} ^{+ 0.21}_{- 0.15} \textrm{ (syst)} ^{+ 0.20}_{-0.14} \textrm{ (theory)}, \non
\mu(\gamma \gamma )_\textrm{CMS} &=&1.564 \textrm{ } ^{+ 0.460}_{-0.419 }.\label{ob1}
\eeq
On the other hand, the other signal strengths are in agreement with SM predictions. 

Motivated by this experimental result, 
there is much literature about  various new physics models  \cite{Carena:2012xa, Chiang:2012qz, Cheon:2012rh, Buckley:2012em,  An:2012vp, Joglekar:2012hb,ArkaniHamed:2012kq,Haba:2012zt,Almeida:2012bq,Abe:2012fb, Delgado:2012sm, Kearney:2012zi, Espinosa:2012in,Dorsner:2012pp,SchmidtHoberg:2012yy,Reece:2012gi,Davoudiasl:2012ig, Carena:2011aa, Cao:2012fz,Carena:2012gp, Ajaib:2012eb, Christensen:2012ei, Hagiwara:2012mg, Benbrik:2012rm,  Wang:2012ts,   Chun:2012jw, Picek:2012ei,  Choi:2012he, SchmidtHoberg:2012ip, Chao:2012xt, Chang:2012ta, Bertuzzo:2012bt, Basso:2012tr, Aoki:2012jj, Basso:2012nh, Funatsu:2013ni, Goertz:2011hj, Carmona:2013cq}. 
Literature about model independent effective operators  is also reported \cite{Berg:2012cg, Grojean:2013kd, Elias-Miro:2013gya}. 
In the Minimal Supersymmetric (SUSY) Standard Model (MSSM) scenario, 
a light stau and a large left-right mixing of staus can appropriately enhance $\mu(\gamma \gamma )$ \cite{Carena:2011aa,Cao:2012fz, Carena:2012gp, Ajaib:2012eb}. However, 
 it was pointed out  that a light stau and a large left-right mixing of staus  may suffer from vacuum instability \cite{CasasLleydaMunoz1996, Rattazzi:1996fb, Hisano:2010re}, 
 and found that the vacuum meta-stability condition severely constrains  the Higgs to diphoton decay rate  \cite{Sato:2012bf, Kitahara:2012pb, Carena:2012mw}.

In order to relax the vacuum meta-stability condition, we have focused on the parameter region where  the mass difference of the two staus is large.
In this parameter region, the heavier  stau raises the quadratic term of the scalar potential, and the vacuum meta-stability condition can be relaxed even if the lighter stau mass $m_{\tilde{\tau}_1}$ is kept light.
Thus, staus with large mass difference may be able to enhance the Higgs to decay rate.
This region has not been considered in the literature \cite{Kitahara:2012pb, Carena:2012mw} yet.
In this paper, we will analyze in detail about this region.

In addition, staus with large mass difference affect the Higgs to $Z \gamma$ decay rate.
Since $SU(2)_L$ isospin and hypercharge differ between the left-handed stau $\tilde{\tau}_L$ and the right-handed stau $\tilde{\tau}_R$, the dependence of the Higgs to $Z \gamma$ decay rate as a function of $\tilde{\tau}_1$  will change  
whether it is dominated by either the left- or right-handed stau.  
Therefore, staus with large mass difference may be able to affect both  the Higgs to diphoton decay rate and the Higgs to $Z \gamma$ decay rate with some correlation  \cite{Zgamma}.

In this paper we analyze the Higgs to diphoton decay rate
in a broad parameter region which includes  stau with large mass difference in the MSSM.
We do not assume any particular high-energy supersymmetry breaking structure. 
In addition, we show that when the mass difference of the two staus is large, the enhancement of the $h \rightarrow \gamma \gamma $ decay decreases monotonically for effective $\mu \tan\beta$, in spite of a relaxation of the vacuum meta-stability condition,
and that the enhancement of the Higgs to diphoton decay rate is up to 40 $\%$ when the lighter stau mass is larger than 100GeV.

This paper is organized as follows.
In section \ref{sec2}, we will evaluate the vacuum transition rate in a broad parameter region which  includes  staus with large mass difference, and  show the effective $\tan\beta$  dependence  of the stability bound. 
In section \ref{sec3}, we will apply the vacuum stability condition which will be calculated in section \ref{sec2} to the Higgs to diphoton decay rate.
Section \ref{sec4} is devoted to our conclusions and discussion.

\section{Vacuum stability}\label{sec2}

First, we consider the scalar potential for the neutral component of the up-type Higgs, $h_u$, the left-handed stau, $\tilde{L}$, and the right-handed stau, $\tilde{\tau }_R$.
Neglecting the potential for the down-type Higgs is a good approximation when  $\tan \beta $ and the CP-odd Higgs mass, $M_A$ are very large. 
The scalar potential can be written as 
\beq
V&=& \frac{1}{2} m_Z^2 \sin ^2\beta ( 1 + \Delta_t )  h_u^2 +\left( m_{\tilde{L}}^2 + \frac{g^2 -g^{\prime 2}}{4} v_2^2 \right) \tilde{L}^2 + \left(  m_{\tilde{\tau}_R}^2 + \frac{g^{\prime 2}}{2} v_2^2 \right) \tilde{\tau}_R^2 \non
& &- \frac{2 m_{\tau} }{v \cos\beta} \frac{1}{1+\Delta _{\tau }} \mu \tilde{L} \tilde{\tau}_R  \left( v_u + \frac{h_u}{\sqrt{2}} \right)  +\frac{g^2 - g^{\prime 2}}{2\sqrt{2}} v_u h_u \tilde{L}^2 + \frac{g^{\prime 2}}{\sqrt{2}} v_u h_u \tilde{\tau}_R^2 \non
& &+  \frac{ m_Z^2 \sin ^2\beta ( 1 + \Delta_t ) }{2 \sqrt{2} v_u} h_u ^3 +\frac{m_Z^2  ( 1 + \Delta_t )}{16 v^2} h_u^4 +\frac{g^2 + g^{\prime 2}}{8} \tilde{L}^4 + \frac{g^{\prime 2}}{2} \tilde{\tau}_R^4 \non
& &+\left\{ \left( \frac{m_{\tau}}{v \cos \beta}\frac{1}{1+\Delta _{\tau }}\right) ^2 - \frac{1}{2} g^{\prime 2}\right\} \tilde{L}^2 \tilde{\tau}_R^2 +\frac{g^2 - g^{\prime 2}}{8} h_u^2 \tilde{L}^2  + \frac{g^{\prime 2}}{4} h_u^2 \tilde{\tau}_R^2, \label{potential}
\eeq
where $H_u^0 = v_u + h_u /\sqrt{2}$, and $v_u = v \sin \beta $ with $v \simeq 174$ GeV. 
$g^{\prime}$ is the gauge coupling for $U(1)_Y$ and $g$ is the gauge coupling for $SU(2)_L$. 
$m_{\tilde{L}} $ and $m_{\tilde{\tau}R} $ are soft SUSY breaking slepton masses.
We take account of only two effects at one-loop order,  $\Delta_t$ and $\Delta_{\tau}$.
Note that the full one-loop effect influences the vacuum stability condition in a few percent \cite{Carena:2012mw}.

The leading log term of the one-loop corrections for top/stop loops $\Delta_t$ is
\beq
\Delta_t \simeq \frac{3}{2 \pi^2}\frac{y_t^4}{g^2 + g^{\prime 2}}\log{\frac{\sqrt{m_{\tilde{t}_1}^2 m_{\tilde{t}_2}^2}}{m_t^2}}, \label{kore}
\eeq
where $y_t  = m_t/v \sin \beta $ and $m_t$ is the weak scale running top quark mass.
When $\Delta _t \simeq 1$, the Higgs boson mass, $m_h^2 \simeq (1 + \Delta _t) m_Z^2 \sin ^2\beta $, is enhanced to 126 GeV.

$\Delta _{\tau }$ is the correction from the tau non-holomorphic Yukawa coupling, and its contribution becomes significant  at large $\tan \beta $ \cite{Carena:1994bv, Pierce:1996zz, Guasch:2001wv}. 
At the tree level, the tau Yukawa coupling is $y_{\tau} = m_{\tau}/ v \cos\beta$.
It is modified at the one-loop level as follows,
\beq
y_{\tau} = \frac{m_{\tau}}{v \cos \beta} \frac{1}{1 + \Delta_{\tau}}.
\eeq
The expression for the $\Delta _{\tau }$ in the  large $\tan \beta$ region is given  as follows, 
\beq
\Delta_{\tau} \simeq -  \mu \tan \beta \left\{ \frac{3}{32 \pi ^2} g^2 M_2 I(m_{\tilde{\nu}\tau} , M_2 , \mu )  - \frac{1}{16 \pi^2 } g^{\prime 2} M_1 I(m_{\tilde{\tau}1 } , m_{\tilde{\tau}2} , M_1)\right\},
\eeq
where
\beq
I(a,b,c) = \frac{a^2 b^2 \ln{(a^2 / b^2 ) } + b^2 c^2 \ln{(b^2 / c^2)} + c^2 a^2 \ln{(c^2/a^2)}}{(a^2 -b^2)(b^2 - c^2)(a^2 - c^2)}.
\eeq
In the above $M_{1(2)}$ is the Bino (Wino) mass. 
If $sign(\mu M_2 ) =+1$,  the sign of $\Delta_{\tau}$ is minus and its absolute value is $\mathcal{O}(0.1)-\mathcal{O}(0.2)$ in the large $\tan \beta$ region. 
Note that since, if $ y_{\tau} \mu$ is held constant, $\Delta _{\tau }$ contribution  raises the scalar potential~(\ref{potential}) through the stau quartic term,  $\tilde{L}^2 \tilde{\tau}_R^2 $ ,  it can relax vacuum stability \cite{Carena:2012mw}. 

Let us define $\tan\beta_{\textrm{eff}}$ as follows,
\beq
\tan\beta_{\textrm{eff}} \equiv \tan \beta \frac{1}{1 + \Delta_{\tau}}.
\eeq
Then, the tau Yukawa coupling is simplified  at large $\tan \beta$,
\beq
y_{\tau} \simeq \frac{m_{\tau}}{v} \tan\beta_{\textrm{eff}}.
\eeq
Since  $\Delta_{\tau}$ is dependent on many SUSY parameters, an analysis becomes complicated. 
For simplicity, in the rest of this paper we treat  $\tan\beta_{\textrm{eff}}$ as an input parameter in the tau sector instead of considering  $\Delta_{\tau} $.

The scalar potential~(\ref{potential}) has some local minima, and the ordinary electroweak-breaking minimum is  $h_u = \tilde{ L} = \tilde{\tau}_R = 0 \GeV$ vacuum.
As the large left-right mixing of staus, i.e. large $\mu \tan\beta_{\textrm{eff}}$,  the new charge-breaking minimum which leads to the $\tilde{L} \neq 0 $ or $\tilde{\tau}_R \neq 0 $ vacuum develops. Then, the minimum becomes lower than the ordinary electroweak-breaking minimum.
Eventually, the electroweak-breaking vacuum (false vacuum) causes vacuum decay to the charge-breaking vacuum (true vacuum) by quantum tunneling effects.

The vacuum transition rate from the false vacuum to the true vacuum can be evaluated by semiclassical technique \cite{Coleman:1977py, Callan:1977pt}. 
The vacuum transition rate per unit space-time volume is evaluated as follows,
\beq
\frac{\Gamma}{V} = A e^{-B},
\eeq
where the prefactor $A$ is the fourth power of the typical energy scale in the potential,  and expected to be roughly $(100{\rm{GeV}})^4$.
$B$ is the Euclidean action which is evaluated on the bounce solution.
When the vacuum transition rate per unit volume $\Gamma / V$ is smaller than the fourth power of the current value of the Hubble parameter $H_0 \simeq 1.5 \times 10^{-42}$ GeV, i.e. the lifetime of the false vacuum is longer than the age of the universe, the false vacuum becomes meta-stable. Then the vacuum meta-stability condition is approximately given as $B \gtrsim  400$\footnote{This meta-stability condition has only logarithmic dependence of the Hubble parameter $H_0$ and the prefactor $A$. So, even if  the value of $H_0$ or  $A$ changes $\mathcal{O}(10) \%$, the change of this condition  is less than one percent.}.

\begin{figure}[tbp]
\begin{center}
\includegraphics[width=120mm]{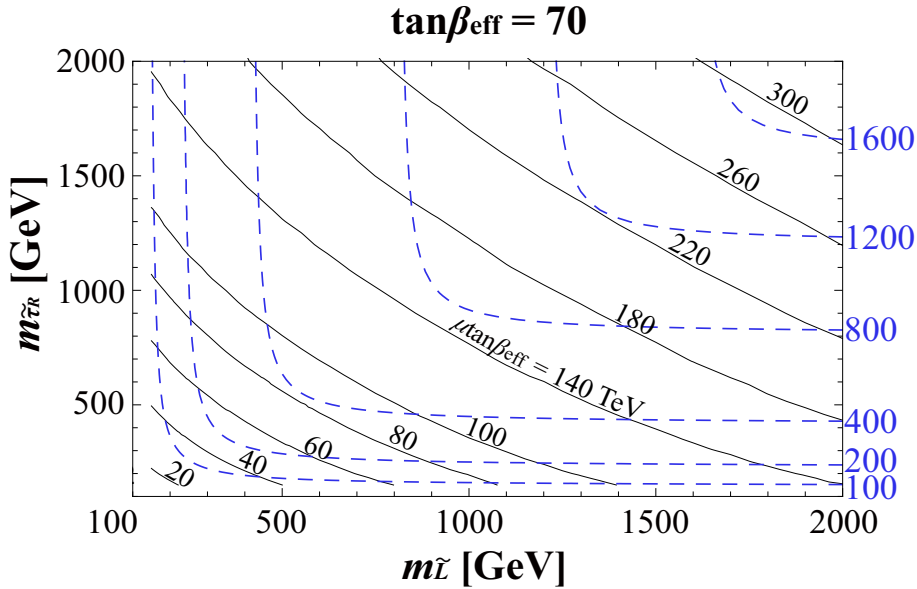}
\end{center}
\caption{The solid lines are contours of the upper bound on $\mu\tan\beta_{\textrm{eff}}$ which  satisfy $B \geq 400$ in $m_{\tilde{L}} - m_{\tilde{\tau }R}$ plane.  The blue dashed lines are contours of the lighter stau mass $m_{\tilde{\tau}1}$ , where $\mu \tan \beta _{{\rm{eff}}}$ is taken to be the  maximum value.
We take  $A_{\tau}= 0 \GeV$, $\tan\beta_{\textrm{eff}} = 70$ and $ m_h = 126\GeV$.}
\label{mutbUB70}
\end{figure}

We analyzed numerically  the  Euclidean action $B$ at zero temperature by {\tt CosmoTransitions 1.0.2}~\cite{CosmoT}, 
which is the numerical package to  analyze  zero and finite temperature cosmological phase transitions for single and multiple scalar fields\footnote{In order to evaluate the bounce  solution and the Euclidean action $B$ numerically,  we applied  modules {\tt tunneling1D.py} and {\tt pathDeformation.py}.  In order to evaluate them at zero temperature, we set the parameter $\alpha = 3$ in the modules.}.
In Figure~\ref{mutbUB70},  we show the upper bounds on $\mu \tan \beta _{{\rm{eff}}}$ which satisfy $B \geq 400$  and contours of the lighter stau mass in $m_{\tilde{L}}$-$m_{\tilde{\tau }_R}$ plane for $A_{\tau } =0$ GeV, $ \tan \beta _{{\rm{eff}}} = 70$, $m_h = 126$ GeV.
The solid lines are the upper bounds on  $\mu \tan \beta _{{\rm{eff}}}$, and
the  blue dashed lines are contours  of the $m_{\tilde{\tau}_1}$, where $\mu \tan \beta _{{\rm{eff}}}$ is taken to be the  maximum value.
When either $m_{\tilde{L}}$ or $m_{\tilde{\tau }_R}$ is large, we find that the vacuum mata-stability condition for $\mu \tan \beta _{{\rm{eff}}}$ is relaxed even if $m_{\tilde{\tau }_1} = \mathcal{O}(100)$ GeV.
Thus, in the case of staus with large mass deference, the upper bound of $\mu \tan \beta _{{\rm{eff}}}$ can be enlarged.

\begin{figure}[tbp]
 \begin{center}
 \includegraphics[width=110mm]{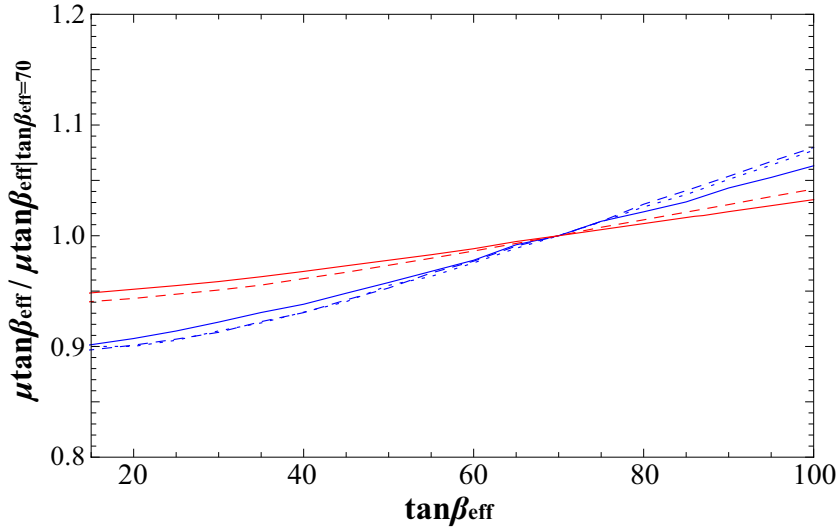}
 \end{center}
 \caption{$\tan \beta _{{\rm{eff}}}$ dependence   of the upper bound on $\mu \tan \beta _{{\rm{eff}}}$  that satisfies $B \geq 400$, normalized by the upper bound on $\mu \tan \beta _{{\rm{eff}}}$ at $\tan\beta_{\textrm{eff}} = 70$. 
 We take  $A_{\tau}= 0 \GeV$ and $ m_h = 126\GeV$.
 The blue solid (dashed, dotted) line corresponds to $m_{\tilde{L}} = m_{\tilde{\tau}R} = 300$ $(1000,\textrm{ }  2000) \GeV$, and the red solid (dashed) line corresponds to $m_{\tilde{L}} =200$ $(300) \GeV, m_{\tilde{\tau}R}  =1500$ $(2000) \GeV $.}
 \label{mutbUB10_100}
\end{figure}

Let us discuss the $\tan \beta _{{\rm{eff}}}$ dependence   of the upper bound on $\mu \tan \beta _{{\rm{eff}}}$.
At large $\tan\beta_{\textrm{eff}}$, the upper bound on $\mu \tan\beta_{\textrm{eff}}$ is alleviated by $\Delta_{\tau}$ effects  \cite{Carena:2012mw}.
In Figure~\ref{mutbUB10_100}, we show the $\tan \beta _{{\rm{eff}}}$ dependence   of the upper bound on $\mu \tan \beta _{{\rm{eff}}}$ for $A_{\tau}= 0 \GeV$,  and $m_h = 126\GeV$, 
normalized by the upper bound on $\mu \tan \beta _{{\rm{eff}}}$ at $\tan\beta_{\textrm{eff}} = 70$. 
The blue lines correspond to the $m_{\tilde{L}} = m_{\tilde{\tau }R}$ case, and  the blue solid (dashed, dotted) line represents $m_{\tilde{L}} = m_{\tilde{\tau}R} = 300$ $(1000,\textrm{ } 2000) \GeV$.
On the other hand, the red lines correspond to the $m_{\tilde{L}} \ll m_{\tilde{\tau }R}$ case, 
 and  the red solid (dashed) line represents $m_{\tilde{L}} =200$ $(300) \GeV, m_{\tilde{\tau}R}  =1500$ $(2000) \GeV $.
 We showed that $\tan\beta_{\textrm{eff}}$ can certainly relax the upper bound of $\mu \tan\beta_{\textrm{eff}}$ that satisfies $B \geq 400$.
 We also found that  the $m_{\tilde{L}} = m_{\tilde{\tau }R}$ cases are more sensitive to $\tan \beta _{{\rm{eff}}}$  than the stau with large mass difference cases.
 This is because, in  the $m_{\tilde{L}} = m_{\tilde{\tau }R}$ cases the charged-breaking vacuum is $\tilde{L } \sim \tilde{\tau}_R \neq 0$, and this vacuum of the scalar potential is more sensitive to $\tan\beta_{\textrm{eff}}$ than the latter cases.  
As a rough estimate, we found that 
when  $\tan\beta_{\textrm{eff}} \sim 90$, the meta-stability condition is relaxed by about $5 \%$, and
when  $\tan\beta_{\textrm{eff}} \sim 20$, the meta-stability condition is tightened by about $10 \%$.

We applied a fit of a function of $m_{\tilde{L}}$ and $m_{\tilde{\tau }_R}$ to  Figure~\ref{mutbUB70}.
The vacuum meta-stability condition  fitting formula is given by
\beq
|\mu \tan \beta_{\textrm{eff}}| < 56.9 \sqrt{m_{\tilde{L}} m_{\tilde{\tau }R}} + 57.1 \left(m_{\tilde{L}}+1.03 m_{\tilde{\tau }R} \right)   - 1.28 \times 10^4 \GeV \non
+\frac{1.67 \times 10^6 \GeV ^2 }{m_{\tilde{L}}+m_{\tilde{\tau }R} }  - 6.41 \times 10^7 \GeV ^3 \left ( \frac{1}{m_{\tilde{L}}^2  } + \frac{0.983}{m_{\tilde{\tau }R}^2}  \right) . \label{metabound}
\eeq
This formula can be applied in the region where  $m_{\tilde{L}}$, $m_{\tilde{\tau }_R} \leq 2$ TeV, the error   of this fit being  less than 1 \% in this region.
It is valid in the case considering not only light staus  but also staus with large mass deference.
Note that an asymmetry of coefficients of $m_{\tilde{L}}$ and $m_{\tilde{\tau }R}$  is reflected by D-term potential, i.e. $g^{\prime 2}$ and $g^2 $ terms in Eq.~(\ref{potential}).

\begin{figure}[tbp]
\begin{center}
\includegraphics[width=120mm]{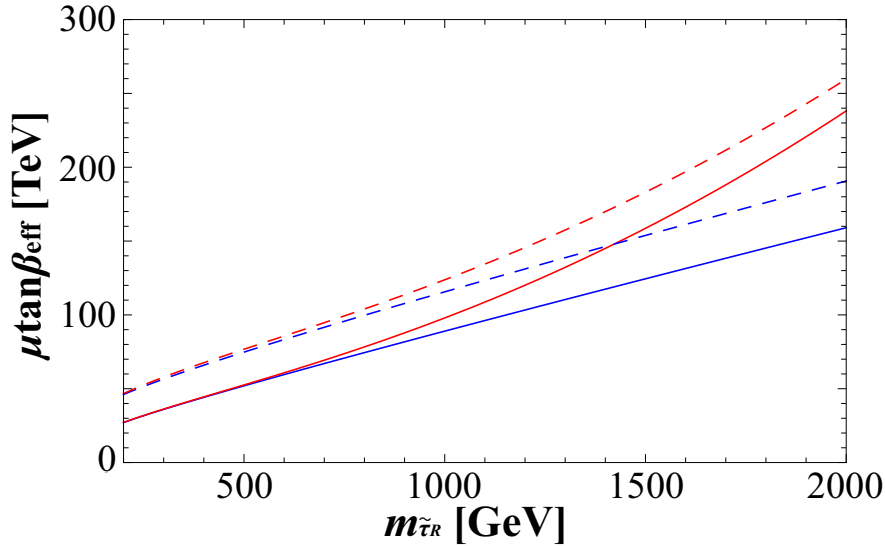}
\end{center}
\caption{Meta-stability bound on $\mu \tan\beta_{\textrm{eff}}$ as a function of $m_{\tilde{\tau}R}$. 
The blue lines correspond the meta-stability condition (\ref{metabound}), and the red lines correspond to the meta-stability bound from Ref.~\cite{Hisano:2010re}. 
We take $m_{\tilde{L} } = 250 \GeV$ (solid) and  $m_{\tilde{L} } = 500 \GeV$  (dashed).}
\label{comperehisano}
\end{figure}

Before proceeding, let us compare  our results to the results of Ref.~\cite{Hisano:2010re}.
The meta-stability condition of the stau sector was  evaluated numerically in the literature\footnote{The result of the vacuum meta-stability conditionin \cite{Hisano:2010re} was obtained as:
\beq
|\mu \tan \beta | &<& 213.5 \sqrt{m_{\tilde{L}} m_{\tilde{\tau }_R} } - 17.0 (m_{\tilde{L}}+ m_{\tilde{\tau }_R})+4.52 \times 10^{-2} \textrm{ GeV}^{-1} (m_{\tilde{L}} - m_{\tilde{\tau }_R} )^2  \non
& &-1.30 \times 10^4 \textrm{ GeV}. 
\label{hisano2}
\eeq}.
Its formula is valid  in the region where both of the staus are light, $m_{\tilde{L}}$, $m_{\tilde{\tau }_R} \leq 600$ GeV. 
We checked that our results reproduce the results of the literature  in this region well, see Figure~\ref{comperehisano}.
We show the meta-stability bound on $\mu \tan\beta_{\textrm{eff}}$ as a function of $m_{\tilde{\tau}R}$. 
The blue lines correspond the meta-stability condition (\ref{metabound}), and the red lines correspond meta-stability bound from Ref.~\cite{Hisano:2010re}. 
We take $m_{\tilde{L} } = 250 \GeV$ (solid) and  $m_{\tilde{L} } = 500 \GeV$  (dashed).
We found that when either $m_{\tilde{L}}$ or $m_{\tilde{\tau }_R}$ is $\mathcal{O}(1)$ TeV, our results deviate from the results given in the literature.
We will discuss the influence of this deviation for the Higgs to diphoton decay rate in next section.

\section{Numerical analysis}\label{sec3}

In this section, we  analyze numerically the ratio of $\Gamma (h \rightarrow \gamma \gamma ) $ to its SM prediction, $\Gamma (h \rightarrow \gamma \gamma ) / \Gamma (h \rightarrow \gamma \gamma )_{SM}$, including the parameter region of staus with large mass difference,
and apply the vacuum meta-stability condition calculated in section~\ref{sec2} to the decay rate. 

In the MSSM,  an analytic expression for the Higgs to diphoton partial width is given in Refs.~\cite{Shifman:1979eb, HiggsHunter},
\beq
\Gamma (h \rightarrow \gamma\gamma) = \frac{\alpha ^2 m_h^3 }{1024 \pi^3} \left | \frac{g_{hWW}} {m_W^2} A^{h}_1 (\tau _W) +\sum_{f} \frac{2 g_{hff}}{m_f} N_{c,f}Q_f^2 A^h_{\frac{1}{2} }(\tau_f) +  A^{h \gamma \gamma}_{SUSY}\right |^2,\\
A^{h \gamma \gamma}_{\text{SUSY}}=\sum_{\tilde{f}} \frac{ g_{h \tilde{f} \tilde{f}}}{ m_{\tilde{f}}^2 } N_{c, \tilde{f}} Q_{\tilde{f}}^2 A_{0}^h (\tau_{\tilde{f}}) + \sum_{i=1,2} \frac{ 2   g_{h \chi^{+}_{i} \chi^{-}_{i}}}{m_{ \chi^{\pm }_{i} }}A_{\frac{1}{2}}^h(\tau_{\chi^{\pm}_i})+\frac{  g_{h H^{+} H^{-}} }{ m_{H^{\pm}}^2} A_0^h (\tau_{H^{\pm}}),
\eeq
where $\tau_i = m_h^2 / 4 m_i^2$, $m_h$ is the lightest CP-even Higgs mass, $N_{c,i}$ is the number of colors of particle $i$, $Q_{i}$ is the electric charge of particle $i$, and the  loop functions $A_{i}^h(\tau)$  and the Higgs couplings $g_{h i i}$ are   given in Appendix \ref{AppA}.

In the light stau scenario, the ratio of the Higgs to diphoton partial width to its SM prediction $\Gamma (h \rightarrow \gamma \gamma ) / \Gamma (h \rightarrow \gamma \gamma )_{SM}$  can be written in a simple formula as follows,
\beq
\frac{\Gamma (h \rightarrow \gamma \gamma )}{\Gamma (h \rightarrow \gamma \gamma )_{SM}} \simeq \left(1+ \sum_{i=1,2}0.05 \frac{m_\tau \mu \tan \beta_{\textrm{eff}} }{ m_{\tilde{\tau}_i}^2}  x^{\tau _i}_L  x^{\tau _i}_R  \right)^2 ,
\eeq
where the stau mass eigenstates are given by $\tilde{\tau}_i = x^{\tau_i}_L  \tilde{\tau}_L + x^{\tau_i}_R  \tilde{\tau}_R$, $(x^{\tau_i}_L)^2 + (x^{\tau_i}_R)^2=1$. 
This simple formula implies that when the lighter stau mass $m_{\tilde{\tau}1}$ is kept light, a large $\mu \tan \beta_{\textrm{eff}}$ can enhance the Higgs to diphoton decay rate. On the other hand, it also implies that 
 the lighter stau $\tilde{\tau}_1$,  which is dominantly constructed by  right-handed stau $\tilde{\tau}_R$   or left-handed stau $\tilde{L}$ (namely, $x^{\tau_1}_L  \ll x^{\tau_1}_R$ or $x^{\tau_1}_L  \gg x^{\tau_1}_R$), suppresses the Higgs to diphoton decay rate.
 In this paper, let us call such a suppression of the Higgs to diphoton decay rate ``the stau L-R mixing suppression". 

In section \ref{sec2}, we showed that  staus with large mass difference can relax the upper bound of $\mu \tan\beta_{\textrm{eff}}$. 
Therefore,  it may be possible to enhance the Higgs to diphoton decay rate in spite of the stau L-R mixing suppression. 
In order to investigate the enhancement of the Higgs to diphoton decay rate when the mass difference of the two staus is large, we apply the vacuum meta-stability condition calculated in section \ref{sec2} to  $\Gamma (h \rightarrow \gamma \gamma ) / \Gamma (h \rightarrow \gamma \gamma )_{SM}$, including the parameter region of staus with large mass difference. 
The Higgs to diphoton decay rate in this region has not been considered  yet \cite{Kitahara:2012pb, Carena:2012mw}.

\begin{figure}[tb]
 \begin{center}
 \includegraphics[width=120mm]{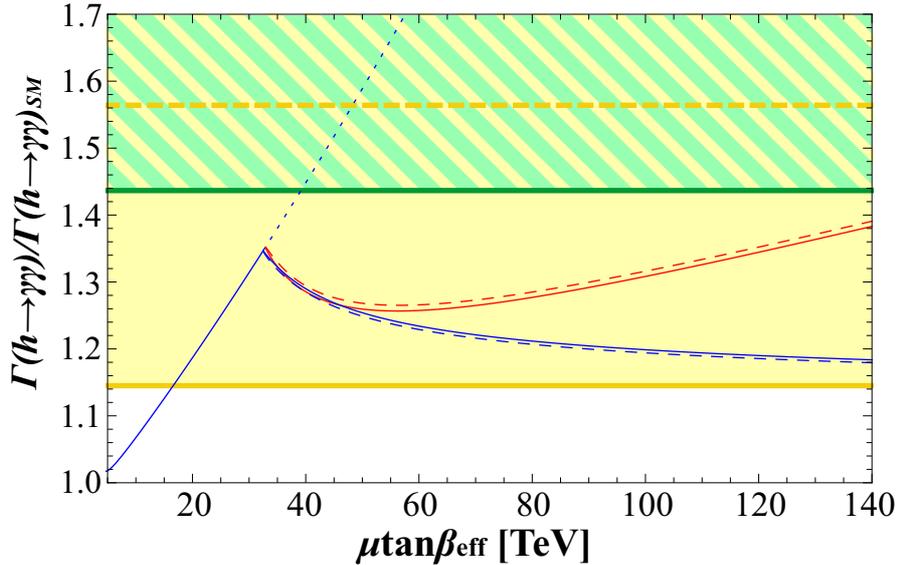}
 \end{center}
 \caption{The upper bound on  $\Gamma (h \rightarrow \gamma \gamma ) / \Gamma (h \rightarrow \gamma \gamma )_{\text{SM}}$ as a function of $\mu \tan\beta_{\textrm{eff}}$, when  $m_{\tilde{L}}$ and $m_{\tilde{\tau}R} $ are varied while keeping $m_{\tilde{\tau}1} = 100\GeV$. 
 We set $\tan \beta_{\textrm{eff}} = 70$.
 The blue solid (dashed) line represents  the upper bound on $\Gamma (h \rightarrow \gamma \gamma ) / \Gamma (h \rightarrow \gamma \gamma )_{SM}$ under the  vacuum meta-stability condition (\ref{metabound}) in the case of $m_{\tilde{L}} \leq  m_{\tilde{\tau }R} \textrm{ } (m_{\tilde{L}} \geq m_{\tilde{\tau }R}  )$ . The red line represents  the bound under the  condition (\ref{hisano2}) which is given by~\cite{Hisano:2010re}.  If we do not consider the vacuum stability, the upper bound of  $\Gamma (h \rightarrow \gamma \gamma ) / \Gamma (h \rightarrow \gamma \gamma )_{\text{SM}}$ is given by the blue dotted line.}
 \label{largemassdiff}
\end{figure}

\begin{figure}[tb]
 \begin{center}
 \includegraphics[width=120mm]{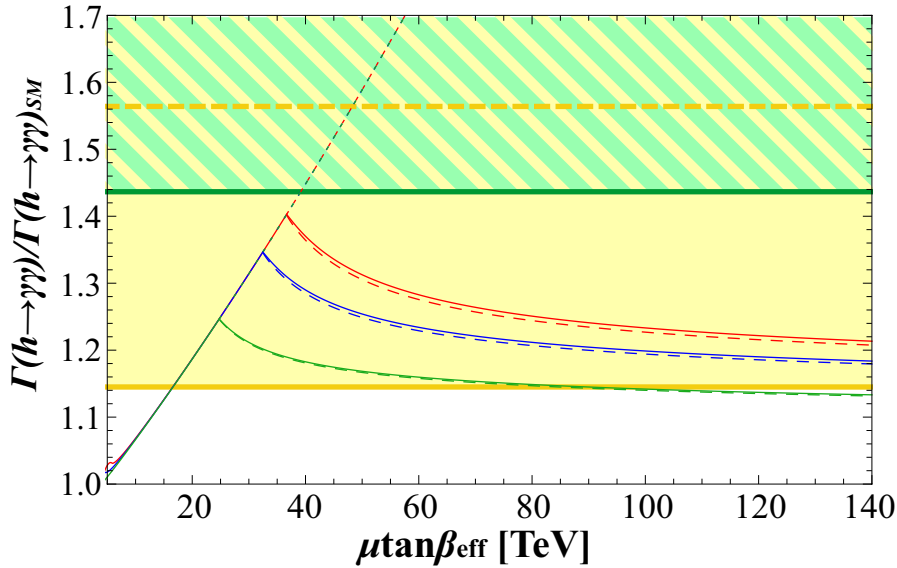}
 \end{center}
 \caption{The upper bound on  $\Gamma (h \rightarrow \gamma \gamma ) / \Gamma (h \rightarrow \gamma \gamma )_{\text{SM}}$ as a function of $\mu \tan\beta_{\textrm{eff}}$, when  $m_{\tilde{L}}$ and $m_{\tilde{\tau}R} $ are varied while keeping $m_{\tilde{\tau}1} = 100\GeV$.
 We set $\tan \beta_{\textrm{eff}} = 70$.
The red and green lines represent the upper bound under a $5\%$-relaxed and a $10\%$-severe vacuum stability condition.
All solid (dashed) lines represent the results in the case of $m_{\tilde{L}} \leq  m_{\tilde{\tau }R} \textrm{ } (m_{\tilde{L}} \geq m_{\tilde{\tau }R}  )$.
If we would not consider the vacuum stability, the upper bound of  $\Gamma (h \rightarrow \gamma \gamma ) / \Gamma (h \rightarrow \gamma \gamma )_{\text{SM}}$ would be given by the dotted line.}
 \label{largemassdiff_relax}
\end{figure}

The result of our numerical computations are drawn in Figure~\ref{largemassdiff} and Figure~\ref{largemassdiff_relax}. 
In Figure~\ref{largemassdiff}, we show the upper bound on $\Gamma (h \rightarrow \gamma \gamma ) / \Gamma (h \rightarrow \gamma \gamma )_{\text{SM}}$ as a function of $\mu \tan\beta_{\textrm{eff}}$, when $m_{\tilde{L}}$ and $m_{\tilde{\tau}R} $ are varied while keeping  $m_{\tilde{\tau}1} = 100\GeV$\footnote{The current lower bound of the stau mass at $95 $ \% CL is obtained by the DELPHI experiment at LEP, 
$m_{\tilde{\tau}1} > 81.9 \textrm{ GeV} $ \cite{Abdallah:2003xe, Beringer:1900zz}. 
We take $m_{\tilde{\tau}1} = 100 \textrm{ GeV} $ as reference. When $m_{\tilde{\tau}1} $ is lighter, the upper bound on  $\Gamma (h \rightarrow \gamma \gamma ) / \Gamma (h \rightarrow \gamma \gamma )_{\text{SM}}$ varies by $\mathcal{O}(10) $ \% }.  
We set  $\tan \beta_{\textrm{eff}} = 70$,  $A_{\tau}=0\GeV$, $m_{\tilde{Q}3} = m_{\tilde{t}R} = m_{\tilde{b}R} = 2 \TeV$, $A_{t} = 3.8 \TeV$, $M_2 = 500\GeV$, $M_3 = 1.5 \TeV$ and $M_A = 10 \TeV$\footnote{ This parameter region gives $m_h \sim 126\GeV$. 
Moreover,  large $M_A$ ensures the validity of analyzing the vacuum meta-stability of the scalar potential in three scalar field spaces ($h_u $, $\tilde{L}$ and $\tilde{\tau}_R$) \cite{Carena:2012mw}. }.
The  light green (yellow) region represents the $1 \sigma$ band of the observational result of   the ATLAS (CMS) collaboration\footnote{Note that the signal strength which was observed by ATLAS (CMS)  is not equivalent to the ratio of the Higgs  partial width to its SM prediction.
For simplicity, we assume $\sigma (pp\rightarrow h)/ \sigma (pp\rightarrow h)_{SM}  \times \Gamma (h \rightarrow \textrm{All})_{SM} / \Gamma (h \rightarrow \textrm{All}) = 1$ in this paper, and the signal strength becomes equivalent to the ratio of the Higgs  partial width to its SM prediction, $\mu (\gamma \gamma)  = \Gamma (h \rightarrow \gamma \gamma)/ \Gamma (h \rightarrow \gamma \gamma)_{SM} $.}.
The blue solid (dashed) line represents  the upper bound on  $\Gamma (h \rightarrow \gamma \gamma)/ \Gamma (h \rightarrow \gamma \gamma)_{SM} $ under the  vacuum meta-stability condition (\ref{metabound})  in the case of $m_{\tilde{L}} \leq  m_{\tilde{\tau }R} \textrm{ } (m_{\tilde{L}} \geq m_{\tilde{\tau }R}  )$.
The red line represents the bound under the condition (\ref{hisano2}) which is given by  \cite{Hisano:2010re}.  If we do not consider the vacuum stability, the upper bound of  $\Gamma (h \rightarrow \gamma \gamma ) / \Gamma (h \rightarrow \gamma \gamma )_{\text{SM}}$ is given by the dotted line.

In the low $\mu \tan\beta_{\textrm{eff}}$ region, $\mu \tan \beta_{\textrm{eff}} \lesssim 33 \TeV $, there are no constraints from vacuum meta-stability, and the Higgs to diphoton decay rate is the most enhanced in the case of  $m_{\tilde{L}} \simeq m_{\tilde{\tau}R}$ \cite{Kitahara:2012pb}.
On the other hand, in the high $\mu \tan\beta_{\textrm{eff}}$ region, $\mu \tan \beta_{\textrm{eff}} \gtrsim 33 \TeV $,  it is constrained severely by vacuum meta-stability. 
In this case, the decay rate is most enhanced in the case that the  mass difference of staus is  non-zero, see Figure~2 in Ref.~\cite{Kitahara:2012pb}. 
Note that there are two cases to enhance the decay rate, $m_{\tilde{L}} \leq  m_{\tilde{\tau }R}$ and $m_{\tilde{L}} \geq  m_{\tilde{\tau }R}$. 
When  $\mu \tan\beta_{\textrm{eff}}$ becomes larger, a larger stau mass difference  is required to enhance the decay rate under the vacuum meta-stability condition, e.g. the decay rate is most enhanced in the case of $(m_{\tilde{L}}, m_{\tilde{\tau}R}) = (155 \GeV, 1650 \GeV)  $ and $(1690 \GeV , 153 \GeV)$,
 when $\mu \tan\beta_{\textrm{eff}}=120\TeV$.
As a result, 
we found that the upper bound on  $\Gamma (h \rightarrow \gamma \gamma ) / \Gamma (h \rightarrow \gamma \gamma )_{\text{SM}}$  decreases monotonically for   $\mu \tan\beta_{\textrm{eff}}$ under  the vacuum meta-stability condition (\ref{metabound}). 
This result implies that the effect of the stau L-R mixing suppression in the Higgs to diphoton decay rate is always larger than that of the relaxation of the vacuum meta-stabiliity.
We also found that 
the enhancement of the Higgs to diphoton decay rate is up to  $35 \%$ at $m_{\tilde{L} } \simeq m_{\tilde{\tau}R}$ and  $\mu \tan\beta_{\textrm{eff}} = 32.5\TeV$. 
This result is  consistent with Ref.~\cite{Carena:2012mw}.
In addition, in the high $\mu \tan\beta_{\textrm{eff}}$ region, we found that the upper bound  on  $\Gamma (h \rightarrow \gamma \gamma ) / \Gamma (h \rightarrow \gamma \gamma )_{\text{SM}}$  under (\ref{metabound})   deviates significantly from the result under  (\ref{hisano2}). 
The reason is that the vacuum meta-stability condition (\ref{hisano2}) is a fitted formula for low $m_{\tilde{L}} $ and $m_{\tilde{\tau}R}$ region. 
When the staus experience a large mass difference, it becomes unreasonable to apply  (\ref{hisano2}) as the vacuum meta-stability condition, see Figure~\ref{comperehisano}.

In Figure~\ref{largemassdiff_relax}, we show the upper bound on $\Gamma (h \rightarrow \gamma \gamma ) / \Gamma (h \rightarrow \gamma \gamma )_{\text{SM}}$ in the case we applied a $5 \%$-relaxed or a $10 \%$-severe vacuum meta-stability condition  to  $\Gamma (h \rightarrow \gamma \gamma ) / \Gamma (h \rightarrow \gamma \gamma )_{\text{SM}}$. 
The  $5 \%$-relaxed and  $10 \%$-severe conditions represent  the meta-stability condition under  $\tan\beta_{\textrm{eff}} \sim 90$ and $\tan\beta_{\textrm{eff}} \sim 20$, see section~\ref{sec2}.
We take the same input  parameters as the case of Figure~\ref{largemassdiff}.
The blue lines  are also the same in Figure~\ref{largemassdiff}.
The red and green lines represent  the upper bound on $\Gamma (h \rightarrow \gamma \gamma ) / \Gamma (h \rightarrow \gamma \gamma )_{\text{SM}}$ under the $5\%$-relaxed and the $10\%$-severe vacuum stability condition.
All solid (dashed) lines represent the results in the case of $m_{\tilde{L}} \leq  m_{\tilde{\tau }R} \textrm{ } (m_{\tilde{L}} \geq m_{\tilde{\tau }R}  )$.
The yellow and green regions are the same in Figure~\ref{largemassdiff}.
As a result, we found that when we apply the $10\%$-severe  and  the $5 \%$-relaxed condition to the Higgs to diphoton decay rate at $m_{\tilde{\tau}1} = 100\GeV$,
the  enhancements in the decay rate are up to  $25\%$ and $40\%$, at $\mu \tan\beta_{\textrm{eff}} = 24.8\TeV$ and $36.7\TeV$, respectively.

\section{Conclusions and Discussion}\label{sec4}

The ATLAS and  CMS collaborations discovered  a Higgs-like particle, and the measurements  of the Higgs couplings to SM particles started at the LHC. 
They have also presented  the result of an excess in the $h\rightarrow \gamma \gamma$ decay channel. 
In the MSSM, this situation can be achieved by a light stau and a large left-right mixing of staus scenario. 
However,  this parameter region is severely constrained by vacuum stability.  
When the mass difference between the two staus is large, the   condition can be relaxed,
since the heavier  stau  raises the quadratic term of  the scalar potential.
In addition, the tau non-holomorphic Yukawa coupling also can relax the condition, and its effect can be expressed by  $\tan \beta_{\textrm{eff}}$ in the stau sector.

In this paper, 
we evaluated the vacuum transition rate in a broad parameter region which includes the parameter region of staus with large mass difference.
We  found that staus with large mass difference can relax the vacuum meta-stability condition sufficiently even if the lighter stau mass $m_{\tilde{\tau}_1}$ is kept light. 
Further, we got a fitting formula of  the vacuum meta-stability condition as Eq.~(\ref{metabound}), 
and studied the $\tan\beta_{\textrm{eff}}$  dependence  of the condition.
For example, when $\tan\beta_{\textrm{eff}} \sim 20 $ or $\tan\beta_{\textrm{eff}} \sim90$, the vacuum meta-stability condition changes to a $10\% $-severe one or to a $5 \%$-relaxed one.

Then, we analyzed numerically the ratio of the  Higgs to diphoton decay rate to its SM prediction $\Gamma (h \rightarrow \gamma \gamma ) / \Gamma (h \rightarrow \gamma \gamma )_{\text{SM}}$, including the parameter region of staus with large mass difference. 
This parameter region has not been considered yet.
We found that when the mass difference of the two staus is large,  the $h \rightarrow \gamma \gamma $ decay rate decreases monotonically for $\mu \tan\beta_{\textrm{eff}}$ under the vacuum meta-stability condition.
This result implies that the effect of the stau L-R mixing suppression in the decay rate is always larger than the effect of the relaxation of the vacuum meta-stability condition.
When $\tan\beta_{\textrm{eff}}=70$,
the  enhancement of  $\Gamma (h \rightarrow \gamma \gamma ) / \Gamma (h \rightarrow \gamma \gamma )_{\text{SM}}$ is up to $35 \%$ at $\mu \tan\beta_{\textrm{eff}} = 32.5\TeV$ in the case of $m_{\tilde{L} } \simeq m_{\tilde{\tau}R}$.   This results is  consistent with Ref.~\cite{Carena:2012mw}.
We also found that when we apply  $10\%$-severe vacuum condition (\ref{metabound}), and  $5 \%$-relaxed one to the decay rate at $m_{\tilde{\tau}1} = 100\GeV$,
 the enhancement of the $\Gamma (h \rightarrow \gamma \gamma ) / \Gamma (h \rightarrow \gamma \gamma )_{SM}$  is up to $25\%$, and $40\%$ at $\mu \tan\beta_{\textrm{eff}} = 24.8\TeV$, and $36.7\TeV$, respectively.
As a result, it is found that an $\mathcal{O}(70)$\% enhancement  of $\Gamma(h\rightarrow \gamma \gamma )/\Gamma(h\rightarrow \gamma \gamma )_{\rm{SM}}$  is difficult in the  light stau scenario in the MSSM. 

Furthermore, staus with large mass difference may be able to affect the Higgs to $Z \gamma$ decay rate \cite{Zgamma}.
Since $SU(2)_L$ isospin and hypercharge differ between the left-handed stau $\tilde{\tau}_L$ and the right-handed stau $\tilde{\tau}_R$,  the dependence of the Higgs to $Z \gamma$ decay rate as a function of  $\tilde{\tau}_1$ will change whether it  is dominated by the left- or right-handed stau.  
If it turns out that the $Z \gamma $ signal strength is not in agreement with the SM prediction by  future experiments, the light stau scenario with a large mass difference might be important.

\section*{Acknowledgements}
We are grateful to Martin Stoll for a careful reading of this paper, and Motoi Endo for useful comments and discussions.
The work of T.K. was supported by  Global COE Program ``the Physical Sciences Frontier", MEXT, Japan. 
The work of T.Y. was supported by an Advanced Leading Graduate Course for Photon Science grant.

\section*{Appendix}
\appendix
\section{Loop functions and Higgs couplings}\label{AppA}
\subsection{The Loop functions $A_{i}^h (\tau)$} 
\beq
A_1^h (\tau) &=& 2 + 3 \tau + 3 \tau (2- \tau )  f(\tau) ,\non
A_{\frac{1}{2}}^h (\tau )&=& -2\tau  \left( 1+ (1 - \tau ) f(\tau ) \right) ,\non 
A_{0}^h(\tau )&=& \tau (1 - \tau f(\tau )),
\eeq
where
\beq
f(\tau) &=&\left\{
\begin{array}{l}
\arcsin ^2 (\sqrt{\frac{1}{\tau}}) ,\textrm{\hspace{52pt} if $\tau \geq 1$,}\\
-\frac{1}{4} \left( \ln{ (\frac{\eta_{+}}{\eta_{-}})} - i \pi \right)^2,\textrm{\hspace{20pt}if $\tau \leq 1$,}
\end{array}
\right. \\
\eta_{\pm} &\equiv& (1 \pm \sqrt{1 - \tau}).
\eeq

\subsection{The Higgs couplings in the MSSM}
 \beq
 g_{hWW}&=& \frac{g^2 v}{\sqrt{2}} \sin (\beta - \alpha) ,\\
 g_{hff(\textrm{up type})}&=& \frac{m_f}{\sqrt{2} v}\frac{\cos \alpha}{\sin \beta},\\
 g_{hff(\textrm{down type})}&=& \frac{m_f}{\sqrt{2} v}\frac{- \sin \alpha}{\cos \beta}, \\
 g_{h \tilde{f}_i  \tilde{f}_i (\textrm{up type})}&=& \left( (- I_{3,L} + (I_{3,L} + Y_L )\sin^2 \theta_W  ) \frac{g m_Z}{\cos \theta_W} \sin (\alpha + \beta )  + \frac{\sqrt{2} m_{f}^2 }{v} \frac{\cos \alpha}{\sin \beta} \right) (x^{f _i}_{L} )^2 \non
 & & + \left( - Y_R \sin^2 \theta_W   \frac{g m_Z}{\cos \theta_W}   \sin (\alpha + \beta )  + \frac{\sqrt{2} m_{f}^2 }{v} \frac{\cos \alpha}{\sin \beta} \right)  (x^{f _i}_{R} )^2 \non   
& & + \frac{\sqrt{2} m_{f}}{ v} \frac{\mu \sin \alpha + A_{f} \cos \alpha }{\sin \beta} x^{f _i}_{L} x^{f _i}_{R}   , \\
g_{h \tilde{f}_i  \tilde{f}_i (\textrm{down type})}&=& \left( (- I_{3,L} + (I_{3,L} + Y_L )\sin^2 \theta_W  ) \frac{g m_Z}{\cos \theta_W} \sin (\alpha + \beta )  - \frac{\sqrt{2} m_{f}^2 }{v} \frac{\sin \alpha}{\cos \beta} \right) (x^{f _i}_{L} )^2 \non
 & & + \left( - Y_R \sin^2 \theta_W   \frac{g m_Z}{\cos \theta_W}   \sin (\alpha + \beta )  - \frac{\sqrt{2} m_{f}^2 }{v} \frac{\sin \alpha}{\cos \beta} \right)  (x^{f _i}_{R} )^2 \non   
& & - \frac{\sqrt{2} m_{f}}{ v} \frac{\mu \cos \alpha + A_{f} \sin \alpha }{\cos \beta} x^{f _i}_{L} x^{f _i}_{R}   , 
\eeq
\beq
g_{h \chi^{+}_{i} \chi^{-}_{i}}&=& \frac{g }{\sqrt{2}} \left(-  \boldsymbol{V}_{i1} \boldsymbol{U}_{i2} \sin \alpha + \boldsymbol{V}_{i2 }\boldsymbol{U}_{i1} \cos \alpha \right) ,\\
g_{h H^{+} H^{-}}&=& g \left( m_W \sin (\beta - \alpha ) + \frac{m_Z \cos 2 \beta }{2 \cos \theta_W } \sin (\alpha + \beta ) \right),
 \eeq
 where $Y_{L/R}$ and $I_{3,L/R}$ represent hypercharge and isospin of left/right-handed sfermion, sfermion mass eigenstates represented by  $\tilde{f}_i = x^{f_i}_L  \tilde{f}_L + x^{f_i}_R  \tilde{f}_R$,  $\theta_W$ is the Weinberg angle, and  $\alpha$ is a rotation angle which translates the gauge-eigenstate basis of the CP-even Higgs mass matrix into the mass-eigenstate basis of one.  The chargino mass matrix can be diagonalized to a real positive diagonal mass matrix by two $2 \times 2$ unitary matrices $\boldsymbol{U}$ and $\boldsymbol{V}$ as follows,
 \beq
 \boldsymbol{U}^{\ast} \begin{pmatrix} M_2 &\sqrt{2}m_W \sin \beta \\ \sqrt{2} m_W \cos \beta & \mu \end{pmatrix} \boldsymbol{V}^{\dag} = \begin{pmatrix} m_{ \chi^{\pm }_{1}} &0 \\ 0 &  m_{ \chi^{\pm }_{2}} \end{pmatrix} .
 \eeq

\bibliography{ref}
 \end{document}